\documentclass[12pt]{article}

\usepackage[english]{babel}

\usepackage[letterpaper,top=2cm,bottom=2cm,left=2.5cm,right=2.5cm,marginparwidth=1.75cm]{geometry}

\usepackage{graphicx}
\usepackage[colorlinks=true, allcolors=blue]{hyperref}

\usepackage{graphicx}
\usepackage{epstopdf, epsfig}
\usepackage{hyperref}
\usepackage{amsmath}
\usepackage{amssymb}
\usepackage{draftwatermark}
\SetWatermarkText{}

\usepackage{svg}
\usepackage{caption}
\usepackage{subcaption}

\usepackage{tikz}

\usepackage{siunitx}
\usepackage{graphicx}
\usepackage{lipsum}
\usepackage{longtable}
\usepackage{pdflscape}

\usepackage{pgf}
\pgfkeys{/pgf/number format/.cd ,precision=2,sci generic={exponent={\times 10^{#1}}}}
\newcommand\convert[1]{\pgfmathprintnumber{#1}}

\usepackage[]{natbib}
\usepackage{filecontents}

\title{\textbf{Scaling relations in quasi-static magnetoconvection with a strong vertical magnetic field}}
\author{Shujaut H. Bader$^1$ and Xiaojue Zhu$^1 \dag$ \\ $^1$\textit{Max Planck Institute for Solar System Research, G\"ottingen, 37077, Germany}. \\ $\dag $ zhux@mps.mpg.de}

\begin{document}
\maketitle
\begin{abstract}
The scaling law for the horizontal length scale $\ell$ relative to the domain height $L$, originating from the linear theory of quasi-static magnetoconvection, $\ell/L \sim Q^{-1/6}$, has been verified through two-dimensional (2D) direct numerical simulation (DNS), particularly at high values of the Chandrasekhar number ($Q$). This relationship remains valid within a specific flow regime characterized by columnar structures aligned with the magnetic field. Expanding upon the $Q$-dependence of the horizontal length scale, we have derived scaling laws for the Nusselt number $Nu$ and the Reynolds number $Re$ as functions of the driving forces (Rayleigh number $Ra$ and $Q$) in quasi-static magnetoconvection influenced by a strong magnetic field. These scaling relations, $Nu \sim Ra/Q$ and $Re \sim Ra Q^{-5/6}$, have been successfully validated using 2D DNS data spanning a wide range of five decades in $Q$, ranging from $10^5$ to $10^9$. The successful validation of the relations at large $Q$ values, combined with our theoretical analysis of dissipation rates and the incorporation of the horizontal length scale's influence on scaling behavior, presents a \textcolor{black}{valid} approach for deriving scaling laws under various conditions.
\end{abstract}

\section{Introduction}\label{sec:intro}

The convective motion of electrically conducting fluids in the presence of an externally imposed magnetic field plays a crucial role in both natural phenomena and practical flows. In astrophysical environments, fluid motions are intricately coupled with magnetic fields, such as in the outer layers of the Sun and other late-type stars. The interaction between thermal convection and magnetic fields in these scenarios has been extensively studied \citep{Proctor_1982, cattaneo_astro_2003, sunMagconSpots}. In industrial systems, magnetic damping is commonly employed in metallurgical applications, while magnetic levitation, pumping, and heating of liquid metal are vital aspects of nuclear engineering \citep{davidsonBook}. Despite its prevalence, understanding the dynamics of magnetoconvection in extreme parameter regimes, particularly in geophysical and astrophysical systems, remains a formidable challenge. Therefore, it is crucial to identify relevant flow regimes and establish corresponding scaling relations to characterize these complex flows. By discerning the points at which flow transitions occur, we can carefully extrapolate scaling laws to a reasonable range of input parameters, thereby providing practical utility in understanding and predicting these phenomena. 

The planar Rayleigh-B\'enard convection (RBC) is a useful model for investigating the fundamental dynamics of magnetic fields on convection. The canonical RBC consists of a fluid layer confined between a pair of plane, parallel boundaries separated by a distance $L$. The boundaries are maintained at a constant temperature difference $\Delta$. In the absence of a magnetic field, the Rayleigh number $Ra={g\beta \Delta L^3}/{\nu\kappa}$ and the Prandtl number $Pr={\nu}/{\kappa}$ characterize the thermal forcing and the ratio of the diffusivities present in the system, where $\beta$ is the coefficient of thermal expansion, $g$ the acceleration due to gravity, and $\nu$ and $\kappa$ being the momentum and thermal diffusivity respectively.  In the presence of an external magnetic field of strength $B_0$, the damping effect by virtue of the Lorentz force must be considered. The strength of the Lorentz force to the viscosity is quantified by the Chandrasekhar number $Q= {B^2_0L^2}/({\rho \nu \mu \eta})$, where $\rho$ is the density, $\mu$ the magnetic permeability, and $\eta$ magnetic diffusivity.

Early studies on magnetoconvection, both experimental as well as theoretical \citep{thompson,chandra,Busse, nakagawa1,jirlow}, are primarily devoted to the linear stability analysis. In recent decades, much focus has been on the scaling relations between the dimensionless heat transport as a function of $Ra$ and $Q$. The dimensionless convective heat transport is characterized by the Nusselt number $Nu$, the ratio of the total heat flux to the conduction heat transfer. In recent decades, experiments by \citep{cioni2000effect,aurnou_olson_2001,burr_mueller,king2013turbulent} have suggested different scalings. \citet{aurnou_olson_2001} found $Nu \sim (Ra/Q)^{1/2}$ in contrast with the findings of \citet{burr_mueller} who report a scaling law $Nu \sim (Ra/Q)^{2/3}$. Both of these studies consider $0 < Q \lesssim 10^4$. \citet{yu2018numerical} conducted numerical simulations over a similar range of values of $Q$, and found the scaling relations between $Nu$ and $Ra/Q$ are in good agreement with the empirical formulas obtained in the experimental studies of \citet{aurnou_olson_2001}. Experimental investigations by \citet{cioni2000effect} pushed the range of Chandrasekhar number by two decades up to $Q \approx 10^6$, and comprehensively characterized two major regimes influenced by the magnetic effects, with an additional transitional regime between them. The first regime, interpreted by them as a condition of marginal stability for the thermal boundary layer, the heat transport scaling $Nu \sim Ra/Q$ holds. For the second major regime, which is characterized by higher thermal forcing, they report $Nu \sim Ra^{0.43}Q^{-0.25}$. Recently, direct numerical simulations of quasi-static magnetoconvection by \citet{yan_calkins_maffei_julien_tobias_marti_2019} with magnetic strengths up to $Q=10^8$ have shown the scaling laws of \citet{aurnou_olson_2001}, \citet{burr_mueller} and \citet{yu2018numerical} to be limited to low values of $Q$. The proposed convective regimes are distinguished by flow characteristics, with the first regime being reminiscent of the linear convection, characterized by laminar, cellular structures. The heat transport scaling law for the second regime, characterized by the existence of quasi-laminar columnar structures, is suggested as $Nu \sim (Ra/Q)^{\gamma}$, where $\gamma \rightarrow 1$ as $Q \rightarrow \infty$. Based on an exponential fit for $\gamma$, a value of $0.95$ is suggested at $Q=10^{16}$. However, as the authors mention, it must be noted that the exponential fit has no physical basis and is shown only to provide a guide for the behavior at large values of $Q$.

Based on the scaling approach introduced by \citet{grossmann2000scaling}, \citet{tillzurner} decomposed the dissipation rates in magnetoconvection into their bulk and boundary layer counterparts and identified four distinct regimes distinguished by the strength of the external magnetic field and the level of turbulence in the flow. Studies by \citet{zurner2020flow, akhmedagaev2020turbulent} report a detailed analysis of the spatial structure of magnetoconvective flows in cylindrical geometry with no-slip walls. Wall modes similar to those in RBC with rotation are observed near the linear stability limit. Furthermore, in these studies, the scaling relations for normalized Nusselt and Reynolds numbers reveal that
the global transport properties approach a universal power law at larger degrees of supercriticality. 

Following \citet{grossmann2000scaling}, \citet{bhattacharyya2006scaling} obtained $Nu \sim Nu(Ra,Q)$ scaling relations for different regimes characterized by weak and strong magnetic fields. It must be noted here \citet{bhattacharyya2006scaling,tillzurner} assumed the length scale associated with the magnetic induction to be the domain height $L$. The presence of a magnetic field opposes fluid motions perpendicular to the field lines while leaving the parallel component unopposed, thus the classical RBC under the effect of an externally applied magnetic field becomes highly anisotropic. Due to anisotropy, the length scale of the flow is significantly altered; the scaling of which remains vital in the scaling laws associated with the heat and momentum transport and Ohmic dissipation. The primary objective of our study is to explore the impact of the horizontal length scale on the scaling relations previously discussed, focusing specifically on the framework of 2D quasi-static assumptions. \textcolor{black}{Although the real-world applications of magnetoconvection are three-dimensional (3D), 2D simulations are substantially cheaper in terms of computational time and thus can be utilized to understand certain aspects of the flow, including scalings. \citet{goluskin2d} and \citet{wang_2d_zonal} are some noteworthy 2D studies that have explored the ability of the convection to drive vertically sheared, large-scale horizontal flow. Building on these works, recently, quasi-2D simulations of magnetoconvection have been used to study the magnetic damping of jet flows which finds applications in probing the mechanisms relevant to damping of large scale azimuthally directed jets on Jupiter \citep{aggarwal2022magnetic}.} Here, our 2D investigation extends beyond previous studies by considering even larger values of $Q$, reaching up to $10^9$. This range of values is particularly relevant for geophysical and astrophysical systems, such as the Earth's outer core, where $Q$ is estimated to be on the order of $10^{16}$ \citep{gillet2010fast}. By identifying scaling relations that converge at values of $Q$ approaching realistic estimates, we aim to enhance our understanding of the fundamental aspects of convection in these systems. 

\section{Physical model and numerical settings}

\subsection{Governing equations}

The dimensionless equations governing the flow of a conducting fluid in Boussinesq approximation, under the action of a quasi-static, vertically imposed external magnetic field $\mathbf{B}= B_0 \mathbf{\widehat{e}_x}$ are,

\begin{subequations} \label{eqn:all_nse_to_j}
\begin{align}
        \frac{\partial \mathbf{u}}{\partial t} + (\mathbf{u} \cdot \nabla )  \mathbf{u} &= -\nabla p + Q\sqrt{\frac{Pr}{Ra}} (\mathbf{J} \times \mathbf{\widehat{e}_x}) + \theta  \mathbf{\widehat{e}_x} + \sqrt{\frac{Pr}{Ra}} \nabla^2 \mathbf{u}, \label{eqn:nse} \\
        \frac{\partial {\theta}}{\partial t} +  (\mathbf{u} \cdot \nabla) \theta &= \frac{1}{\sqrt{RaPr}} \nabla^2 {\theta}, \label{eqn:temp} \\
        \mathbf{J} &= -\nabla \Phi + \mathbf{u} \times \mathbf{\widehat{e}_x}, \label{eqn:currentDen}  \\
        \nabla \cdot \mathbf{u} &= 0, \label{eqn:divg} \\
        \nabla \cdot \mathbf{J} &= 0. \label{eqn:divgofJ}
\end{align}
\end{subequations}

where $\mathbf{u}$, $p$ and $\theta$ represent the velocity, pressure and temperature fields. The vertical unit vector anti-parallel to the direction of gravity is $\mathbf{\widehat{e}_x}$, and $\Phi$ and $\mathbf{J}$ represent the electric potential and the current density respectively. At all places in the paper, $x$ and $y$ represent the vertical and horizontal directions respectively. The input parameters $Ra$, $Pr$, and $Q$ are defined in section \ref{sec:intro}. The rest of the symbols have their usual meanings. In equation \ref{eqn:all_nse_to_j}, domain height $L$, free-fall velocity $u_{ff} = {(\beta g \Delta L)}^{1/2}$, free-fall time scale $t_{ff} = L/u_{ff}$, imposed temperature difference $\Delta$, free-fall velocity-based dynamic pressure $\rho_0 u^2_{ff}$, and the magnitude of the imposed magnetic field $B_0$ are adopted for non-dimensionalization. All lateral boundaries are set to periodic. At the top and bottom walls, stress-free boundary conditions for $\mathbf{u}$, isothermal for $\theta$, and an insulating boundary condition on $\Phi$ are imposed, $u_x = {\partial u_y}/{\partial x} = 0 \ \text{at} \ x = 0,1;
   \quad \theta = 1(0) \text{ at } x = 0(1); \quad {\partial \Phi}/{\partial x} = 0 \ \text{at} \ x=0, 1$.
To compute $\mathbf{J}$, the constraint of charge conservation, given by equation \ref{eqn:divgofJ}, is invoked to obtain a Poisson equation for $\Phi$, $\nabla^2 \Phi = \nabla \cdot (\mathbf{u} \times \mathbf{\widehat{e}_x}).$

The Reynolds number, $Re$, referred to in the subsequent sections is defined as, $Re = {\mathcal{U}L}/{\nu}$, where $\mathcal{U}$
 is the characteristic velocity. Depending on if the characteristic velocity is $u_{rms} = \left(     \overline{\langle u^2_x \rangle_A + \langle u^2_y \rangle_A} \right )^{1/2} $, $u_{x,rms}=\left (     \overline{\langle u^2_x \rangle_A }\right )^{1/2} $, or $u_{y,rms} = \left (\overline{\langle u^2_y \rangle_A }\right )^{1/2} $, we further define $Re$, $Re_x$ and $Re_y$ to characterize the total, vertical and horizontal Reynolds numbers respectively. $\langle \cdot \rangle_A$ and $\overline{(\cdot)}$ denote area-average over the whole 2D domain, and time-averaging respectively.

\subsection{Numerics}
A conservative, second-order centered spatial discretization  is employed on a staggered grid \citep{afidmain}. Time marching is performed with a fractional-step third-order Runge-Kutta (RK3) scheme, in combination with the Crank-Nicholson scheme for the implicit terms. In the following section, the implementation of the subroutines for $\Phi$ and the validation of the code are described in detail.


\subsubsection{\textcolor{black}{Code validation}}
\color{black}
\textcolor{black}{In this work, we have used AFiD \citep{afidmain,zhugpu} }as the base solver, and implemented the subroutines for solving a Poisson equation for the electric potential $\Phi$, and an explicit Lorentz body force term. Temperature and velocities are computed on cell faces, with $\theta$ and $u_x$ located at the same face (to avoid interpolation errors in calculating the term $\theta \mathbf{\widehat{e}_x}$). The scalar variables $p$ and $\Phi$ are computed at the cell centers. Using the cell-centered values of $\Phi$, the components of the current density are then constructed on the cell faces, which are used to finally compute the components of the Lorentz force. 

The solver was validated by comparing the Nusselt number against the values reported in \citet{yan_calkins_maffei_julien_tobias_marti_2019} at diffferent Rayleigh numbers away from the onset. Table \ref{tab:nu_compare_3d} shows the Nusselt number from 3D test runs at $Q=10^6$ to reproduce results from the literature. The resolutions employed for these cases in order, from smallest to largest $Ra$, are: $(96 \times 180 \times 180)$, $(256 \times 384 \times 384)$, and $(512 \times 1024 \times 1024)$, where the first entry corresponds to the grid number in the vertical $(x)$ direction. $Nu_p = \overline{ \langle d \theta/dx \rangle }_{x=0}$ is the plate Nusselt number, $Nu_{\nu + \eta} = 1+ \sqrt{RaPr} \left ( \overline{\langle \epsilon_{\nu} \rangle} +\overline{\langle \epsilon_{\eta} \rangle} \right ) $ and $Nu_{\kappa} = \sqrt{RaPr} \ \overline{\langle \epsilon_{\kappa} \rangle}$ represent the Nusselt numbers calculated from the global, time- and area-averaged viscous $(\overline{\langle \epsilon_{\nu}  \rangle} = \sqrt{Pr/Ra} \  \overline{\langle |\nabla \mathbf{u}|^2 \rangle} )$ and Ohmic dissipation $(\overline{\langle \epsilon_{\eta}  \rangle} = Q\sqrt{Pr/Ra} \  \overline{\langle |\nabla \mathbf{J}|^2 \rangle} )$, and thermal dissipation $(\overline{\langle \epsilon_{\kappa}  \rangle} = \sqrt{Pr/Ra} \  \overline{\langle |\nabla {\theta}|^2 \rangle} )$ respectively. 

\begin{figure}
\centering 
    \includegraphics[width=1\textwidth]{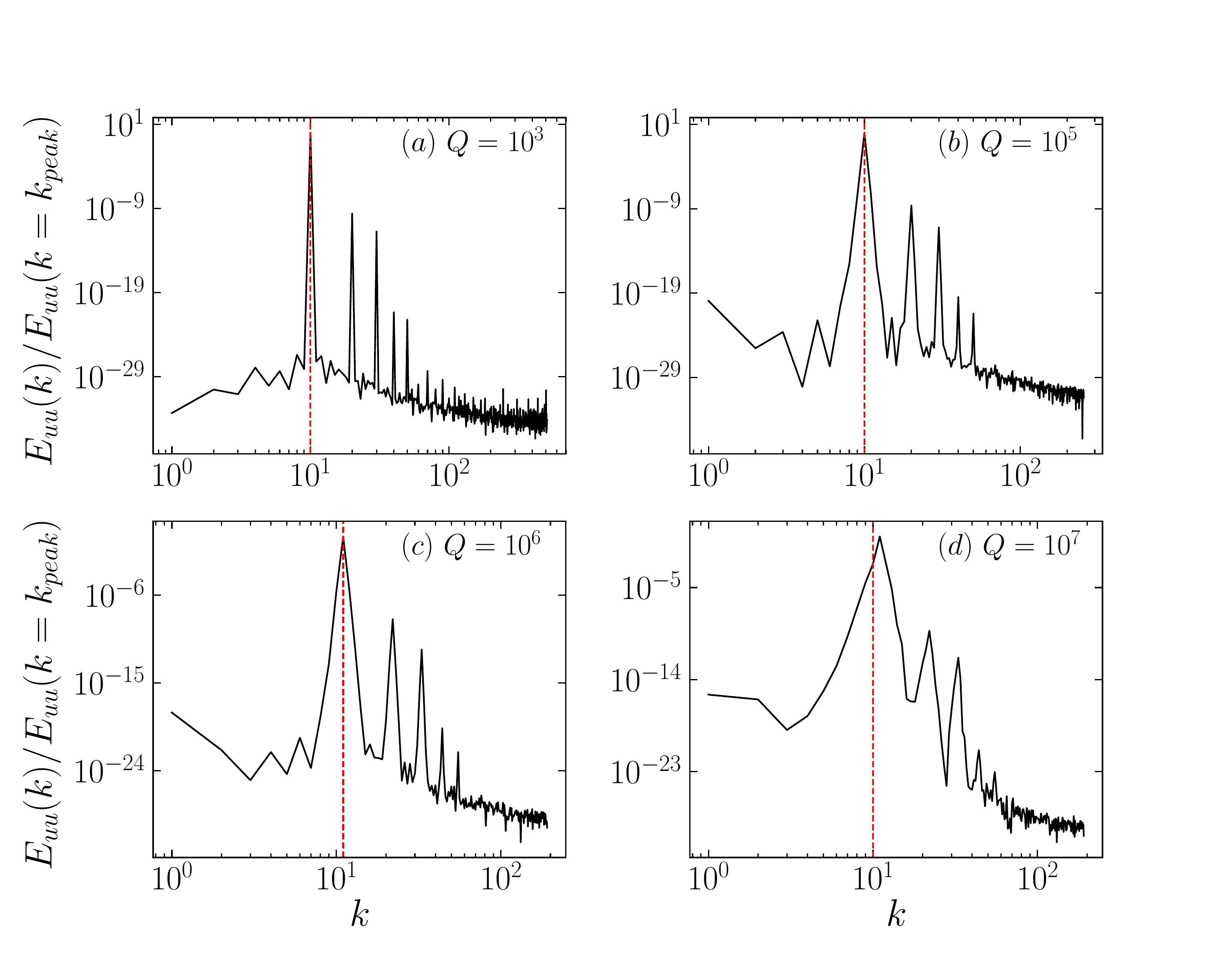}
    \caption{Normalized kinetic energy spectra of the vertical velocity at (a) $Q=10^3, Ra = 1.530 \times 10^4 \gtrapprox Ra_c = 1.521 \times 10^4$, (b) $Q=10^5,  Ra = 1.088 \times 10^6 \gtrapprox Ra_c = 1.078 \times 10^6$ (c) $Q=10^6,  Ra = 1.038 \times 10^7 \gtrapprox Ra_c = 1.028 \times 10^7$, and (d) $Q=10^7,  Ra = 1.016 \times 10^8 \gtrapprox Ra_c = 1.006 \times 10^8$. Vertical dashed red lines show the results from linear theory.} 
    \label{fig:marginal_validation}
\end{figure}

In addition to that, the dominant length scale at $Ra \gtrapprox Ra_c$ (very close to the onset) was also calculated from the energy spectra of the vertical velocity, and validated against the predictions from the linear theory. In figure \ref{fig:marginal_validation}, the normalized kinetic energy spectra of the vertical velocity $E_{uu}(k) = \sum [\widehat{u}_x(k) \widehat{u}^*_x(k) $] computed at mid-depth are shown. $\widehat{u}_x$ and $\widehat{u}^*_x$ are the Fourier transform of $u_x$ and its complex conjugate respectively.  The dominant integer wavenumber represented by the highest peak of spectra, indicates the number of (near-onset) critical horizontal wavelengths ($\lambda_c$) present in the domain. The aspect ratio for $Q=10^3$, $10^5$ and $Q=10^7$ cases was set to $\Gamma = 11.0 $, $\Gamma = 5.0$ and $\Gamma = 2.25$ respectively which admit $\approx 10\lambda_c$. For $Q=10^6$ case, $\Gamma = 3.5$, which allows for $\approx 11\lambda_c$ to be resolved. The number of critical horizontal wavelengths observed in our simulations match reasonably well with the linear theory predictions for a given aspect ratio.  

\begin{table}
    \centering
    \begin{tabular}{c c c c c c c}
        $Ra$ & $Nu_p$ & $Nu_{\nu + \eta}$ & $Nu_{\kappa}$ &  $Nu_{avg}$ & $ Nu $(\citet{yan_calkins_maffei_julien_tobias_marti_2019}) & Error $(\%)$ \\ \hline
         $1.1 \times 10^7$ & 1.140 & 1.140 & 1.140 & 1.140  & 1.135 & 0.44 \\
         $4 \times 10^8$ & 28.471 & 28.473 & 28.505 & 28.483  & 28.580 & 0.34 \\
         $1 \times 10^9$ & 49.766 & 49.704 & 49.769 & 49.746  & 49.70 & 0.09\\
         \hline 
    \end{tabular}
    \caption{Nusselt number from 3D test runs at $Q=10^6$ to reproduce results from the literature.   }
    \label{tab:nu_compare_3d}
\end{table}

\color{black}

\section{Results and discussion}
\subsection{Parameter range and flow structure}
We have considered five sets of simulations with $Q$ ranging from $10^5 - 10^9$. For each set, at a fixed $Q$, the Rayleigh number is varied up to $Ra \gtrsim 100Ra_c$, $Ra_c$ being the critical Rayleigh number at a given $Q$, which is calculated from the expression of the Rayleigh number corresponding to the marginal stability of the horizontal wavenumber. For details, the interested reader is referred to \citet{chandra}. In all of our simulations, we have used $Pr=1$. \textcolor{black}{More details on grid resolution and the parameter ranges explored are listed in Appendix \ref{appA}.}

\begin{figure}
    \centering
    \includegraphics[width=0.7\textwidth]{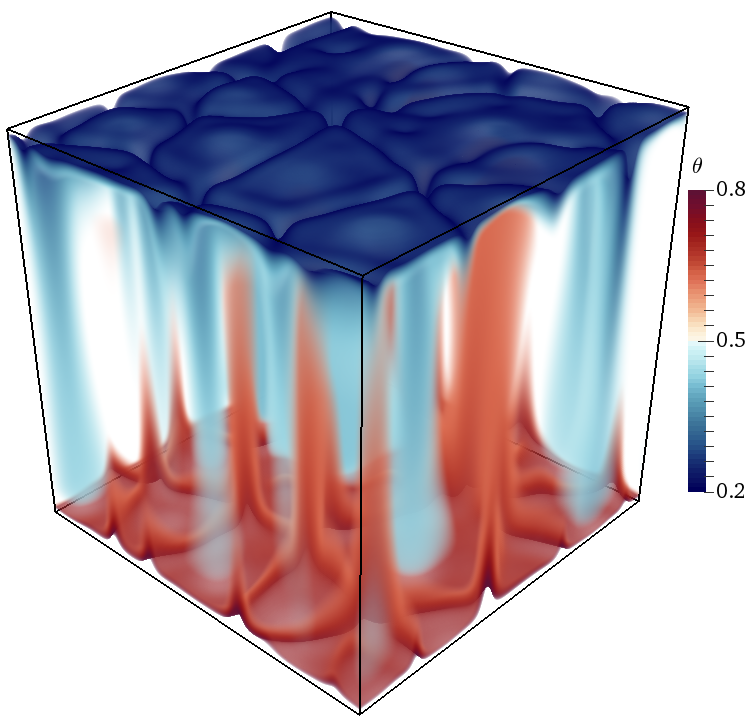}
    \caption{\textcolor{black}{Columnar structures in a 3D simulation at $Ra = 1\times 10^8$, $Q=10^6$, $\Gamma =1.0$. Color represents instantaneous temperature. In 3D, these values lie in the columnar regime.}  }
    \label{fig:3d_colms}
\end{figure}

Figure \ref{fig:colm_contours} shows the visualization of the effect of increasing $Ra$ at a fixed \textcolor{black}{$Q$ ($Q=10^6$)}. The flow structures at all $Ra$ are stabilized by the Lorentz force, resulting in the convection directed along the externally imposed magnetic field. For the cases in the first row from the bottom, flow is dominated by cellular structures which eventually morph into columns as $Ra$ is increased. The transport of heat in the cellular regime is primarily via conduction. In the columnar regime, the length scale of the structures stays more or less the same, until a third regime of magneto-plumes emerges. Upon further increasing the Rayleigh number, the plumes begin to merge, characterized by stronger buoyancy relative to the Lorentz force. We will not be dealing with the scaling laws in this merging-plume regime. With increasing $Ra$, the thermal boundary layer thickness also decreases. This has important consequences on the heat transfer scaling within the columnar regime, which we will discuss later. \textcolor{black}{The flow in cellular, columnar and plume regimes is strictly steady. Unsteadiness kicks in when the stabilizing effect of the Lorentz force is weakened at around $Ra/Ra_c \sim 80$, which also marks the onset of the merging plume regime. Figure \ref{fig:colm_contours}r shows such a case. Upon comparisons with the flow structures in 3D simulations, we observe that the inter-column distance in 2D is identical in the horizontal direction. However, no such uniformity is seen in 3D simulations of \citet{yan_calkins_maffei_julien_tobias_marti_2019}. To illustrate the differences, in Figure \ref{fig:3d_colms}, we have presented instantaneous temperature contours computed in a 3D simulation in the columnar regime. Given insulating boundary conditions $(\partial_x \Phi|_{x=0,1}=0)$ (non-conducting walls), electric currents induced by the horizontal motion $(\mathbf{u} \times \widehat{\mathbf{e}}_x)$ will close within the convective roll itself. In 3D, this gives rise to currents in the horizontal plane $(y-z)$ near the boundaries. In the case of 2D, since one horizontal component of velocity is zero ($u_z$ in the present case), the current associated with that component also ceases to exist, and only one of the components, $u_y$, remains. This gives rise to the Lorentz force which is confined to the \textit{only} horizontal degree of freedom $(y)$ existing in the 2D flow. Since everything is constrained to a single horizontal dimension, there is a tendency of the flow structures to adhere to a specific equidistant pattern, until the stabilizing effect of the Lorentz force is weakened at relatively higher $Ra$. At that point, the interplume distance starts to vary.}

\begin{figure}
\captionsetup{skip=0pt}
     \centering
     \begin{subfigure}[b]{0.85\textwidth}
         \centering
         \includegraphics[width=1\textwidth]{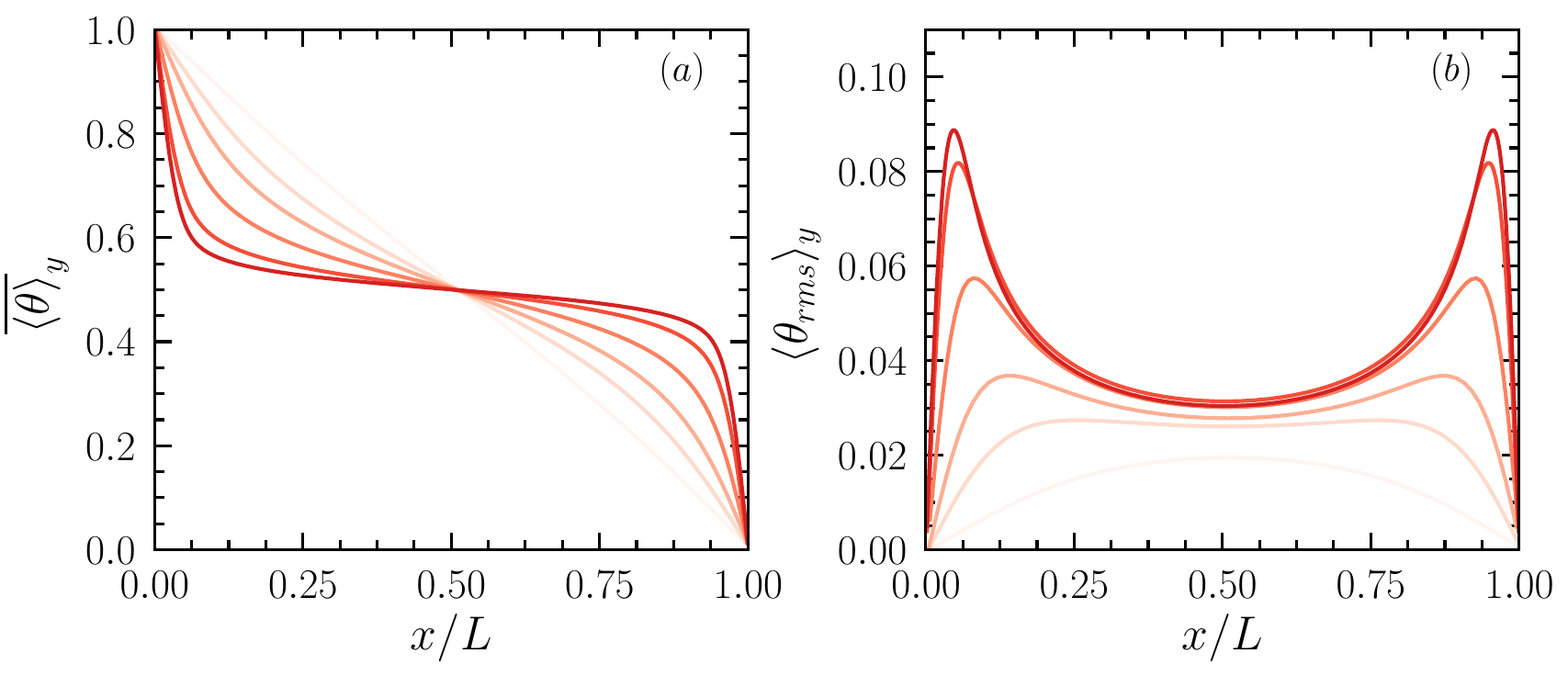}
    \label{fig:Tm_Trms}
     \end{subfigure}
     \begin{subfigure}[b]{0.9\textwidth}
         \centering
         \includegraphics[width=1\textwidth]{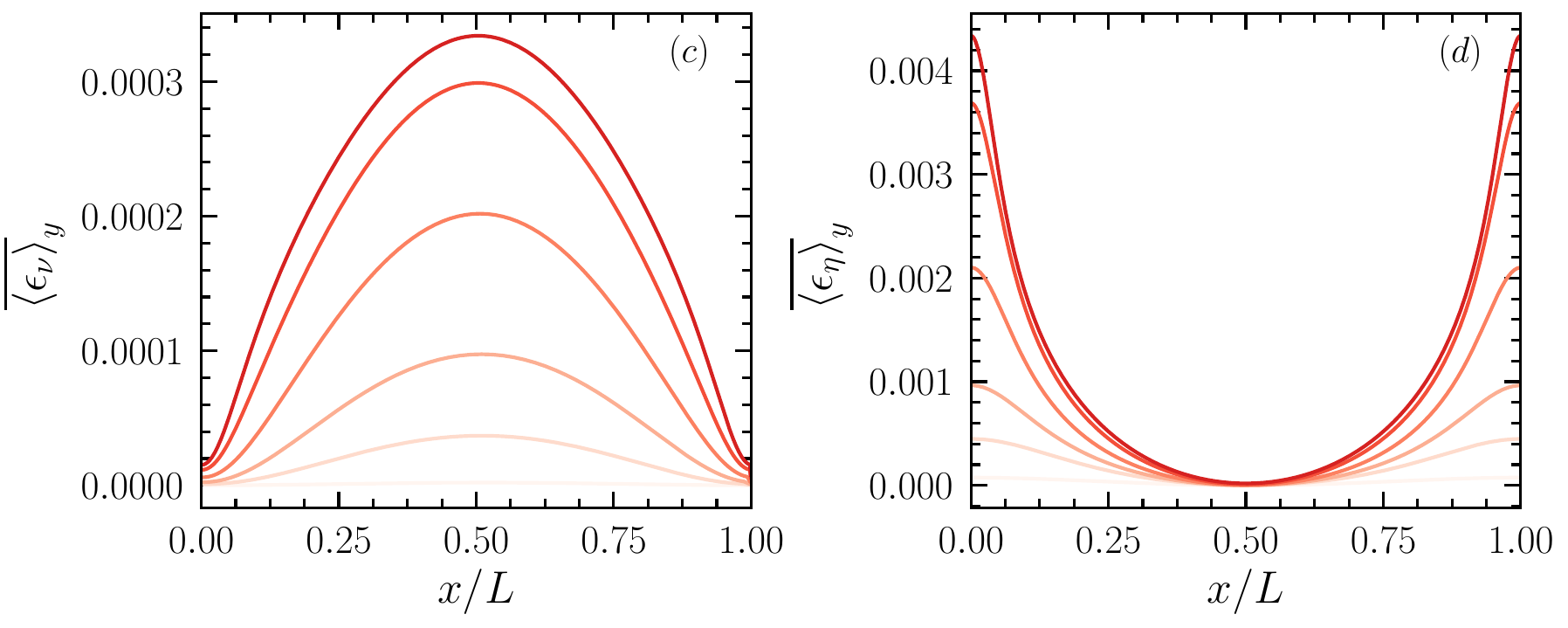}
    \label{fig:dissUB_vert}
     \end{subfigure}
     \begin{subfigure}[b]{0.42\textwidth}
         \centering
         \includegraphics[width=1\textwidth]{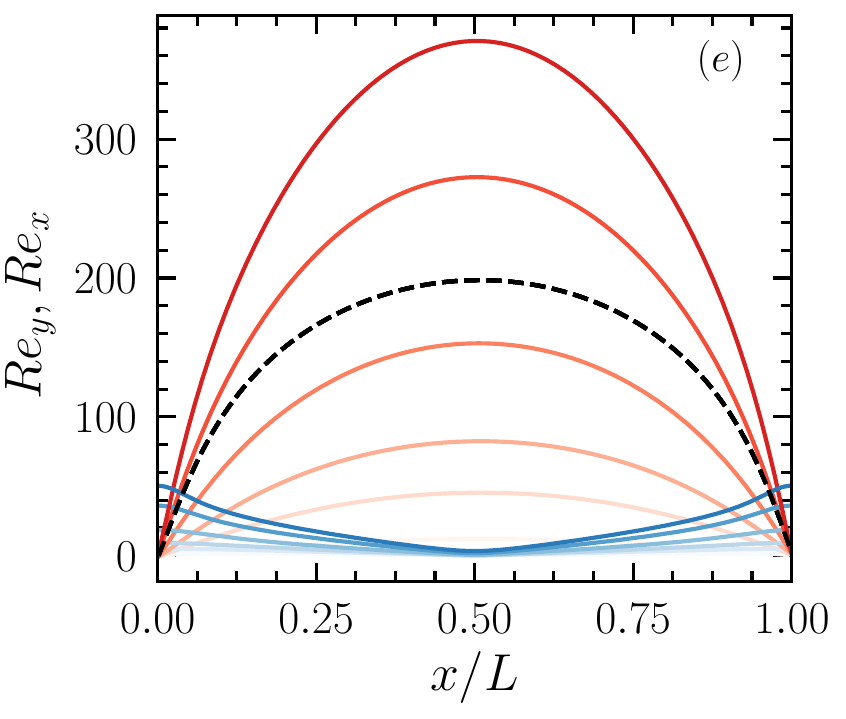}
    \label{fig:ReVReH_vert}
     \end{subfigure}
        \caption{\small Vertical profiles of the (a) time- and horizontally-averaged mean temperature $\overline{\langle \theta \rangle }_y$, and (b) horizontally-averaged rms fluctuations of temperature $\langle \theta_{rms} \rangle_y$, (c) viscous dissipation $
\overline{\langle \epsilon_{\nu} \rangle }_y$, and (b) Ohmic dissipation $\overline{\langle \epsilon_{\eta} \rangle }_y$, and (e) Reynolds number. Blue curves represent $Re_y$, red $Re_x$. Darker curves represent higher $Ra$. The data are plotted for $Q=10^6$. The Rayleigh numbers from smallest to largest are: $1.1\times 10^7, 1.5\times 10^7, 2\times 10^7, 3\times 10^7,5\times 10^7,7\times 10^7$. \textcolor{black}{The dashed black curve represents the vertical Reynolds number ($Re_x$) computed in a 3D simulation at $Ra=7\times 10^7$. The aspect ratio for the 3D simulation is $\Gamma_{3D}=1.0$. However, that is not expected to change the conclusion.} }
        \label{fig:vertical_profiles_all}
\end{figure}

The anisotropy in the convective flow can be characterized by examining the Reynolds number based on the vertical and horizontal components of the rms velocity. Figure \ref{fig:vertical_profiles_all}e clearly shows that the vertical component of the velocity dominates over the horizontal component. As $Ra$ is increased, the vertical component increases significantly relative to the horizontal component.
\begin{figure}
    \centering
     \includegraphics[trim={0cm 0cm 0cm 0cm},clip,width=1\textwidth]
    {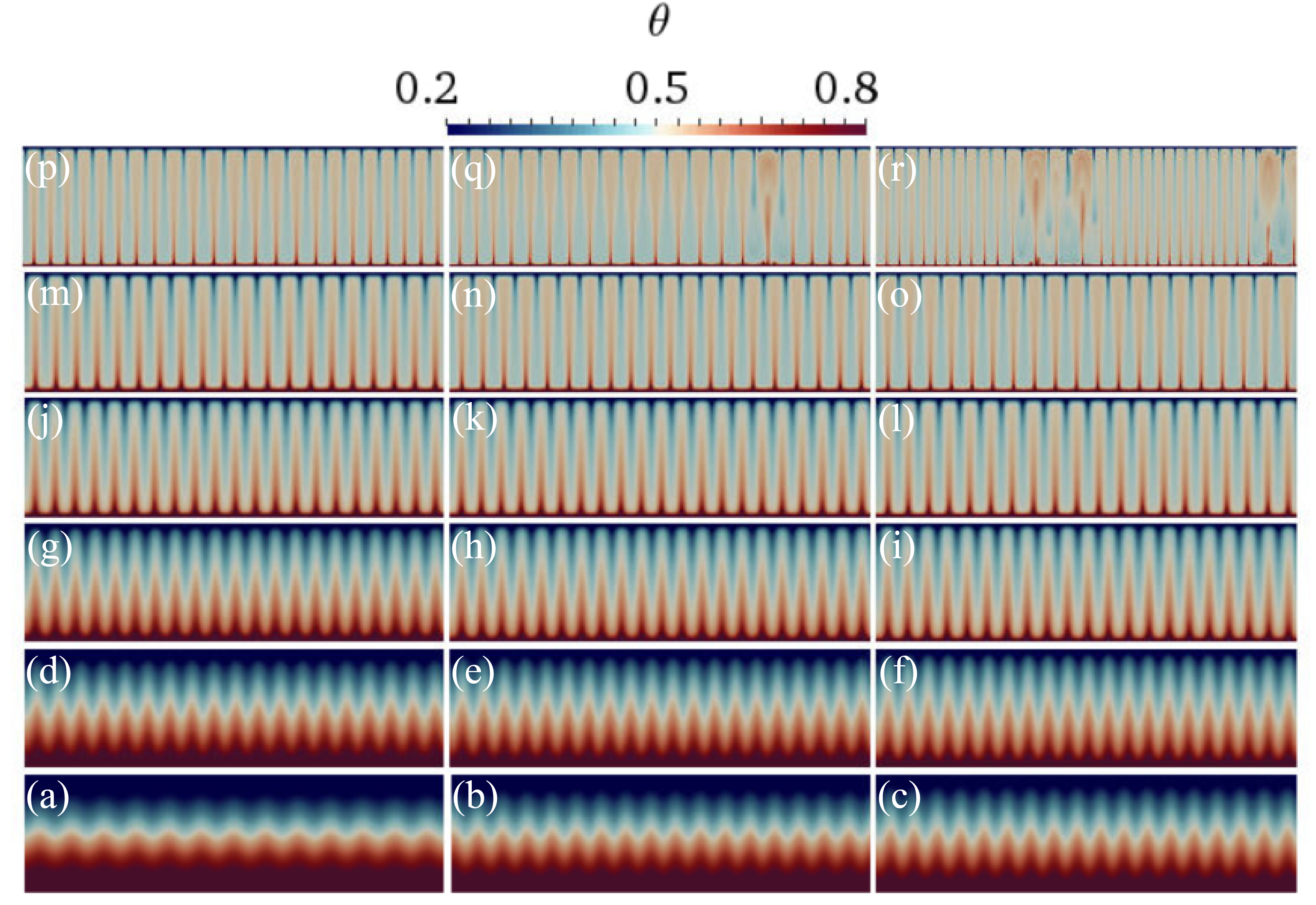}
    \caption{\textcolor{black}{Instantaneous temperature contours in 2D quasi-static magnetoconvection at $Q=10^6$. $Ra$ increases from $(a)$ through $(r)$, as listed in the Appendix for $Q =10^6$, from smallest $(1.1\times10^7)$ to the largest $(8\times10^8)$. From $(a)$ through $(r)$, $Ra/Ra_c =$ $1.07$, $1.26$, $1.46$, $1.65$, $1.94$, $2.43$, $2.92$, $3.89$, $4.86$, $5.84$, $6.81$, $8.75$, $9.73$, $14.59$, $19.45$, $24.32$, $38.91$, $77.82$, $97.28$.  The colormap values have been adjusted to $[0.2,0.8]$ for better visibility.}}
    \label{fig:colm_contours}
\end{figure}

In figures \ref{fig:vertical_profiles_all}a and \ref{fig:vertical_profiles_all}b, vertical profiles of time- and horizontally-averaged mean temperature $\overline{\langle \theta \rangle }_y$, and horizontally-averaged rms fluctuations of temperature $\langle \theta_{rms} \rangle_y$ at $Q=10^6$ are presented. For lower values of $Ra$, the rms fluctuations of temperature are small, and the mean temperature follows the conduction profile. At larger values of $Ra$, the mean temperature tends to be an isothermal profile in the bulk, and the near-wall peaks in the rms profiles indicate the presence of well-developed thermal boundary layers. The darkest shade in figure \ref{fig:vertical_profiles_all} represents $Ra=7\times 10^7$, which lies in the columnar regime (see figure \ref{fig:colm_contours}k).

Figures \ref{fig:vertical_profiles_all}c and \ref{fig:vertical_profiles_all}d show the depth-dependence of the time- and horizontally-averaged viscous $(\epsilon_{\nu} = (Pr/Ra)^{1/2} \langle \overline{|\nabla \mathbf{u}|^2} \rangle_y)$ and Ohmic dissipation $(\epsilon_{\eta} = Q(Pr/Ra)^{1/2} \langle \overline{| \mathbf{J}|^2} \rangle_y)$. Since we have adopted stress-free boundary conditions, the viscous dissipation is larger in the bulk of the domain, away from the boundaries. Ohmic dissipation, on the other hand, is dominant near the boundaries and decreases in the interior of the domain. At larger $Ra$, an increase in the Ohmic dissipation is expected due to larger currents generated by enhanced convection.

\subsection{Scaling analysis}
\begin{figure}
\captionsetup{skip=0pt}
     \centering
     \begin{subfigure}[b]{0.45\textwidth}
         \centering
         \includegraphics[width=1\textwidth]{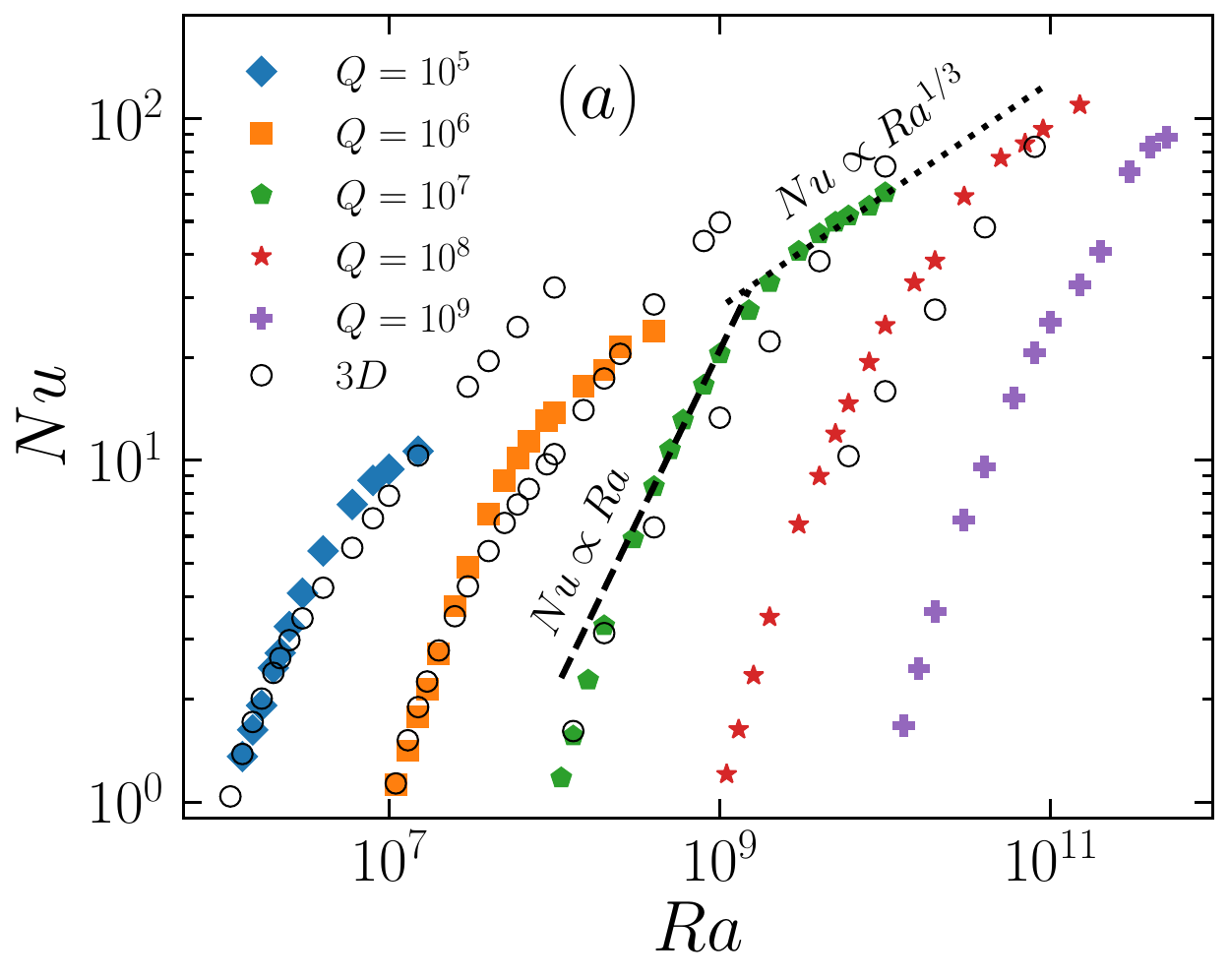}
     \end{subfigure}
     \begin{subfigure}[b]{0.45\textwidth}
         \centering
         \includegraphics[width=1\textwidth]{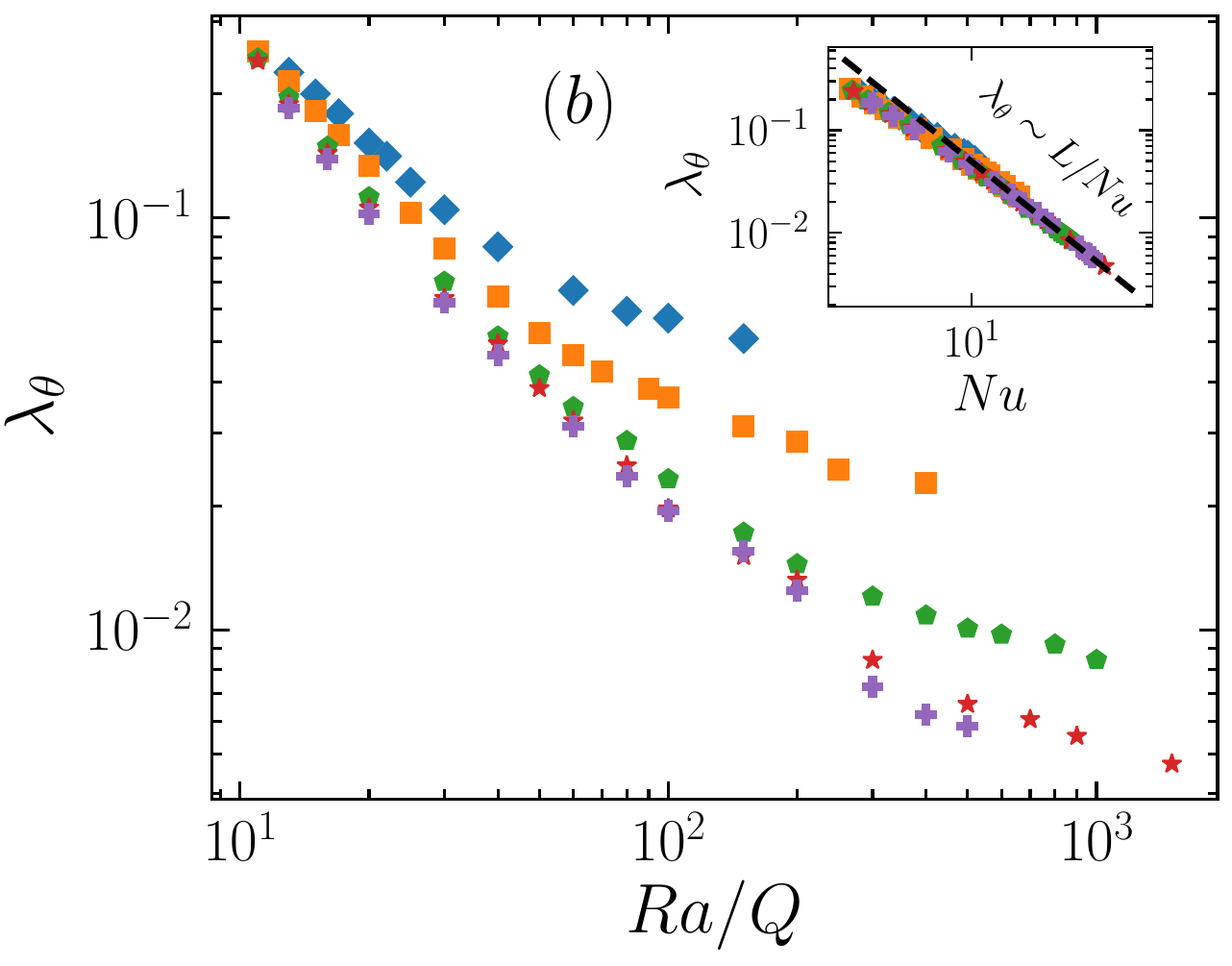}
     \end{subfigure}
        \caption{(a) Nusselt number. Columnar and plume regimes are marked by $Nu \propto Ra$ (dashed) and $Nu \propto Ra^{1/3}$ (dotted) lines for $Q=10^7$,  (b) thermal boundary layer thickness as a function of the supercriticality parameter. 3D data adapted from \citet{yan_calkins_maffei_julien_tobias_marti_2019} are shown for $Q = 10^5, 10^6, 10^7$ and $10^8$. }
        \label{fig:nuss_and_lambda_both}
\end{figure}

\begin{figure}
    \centering
\includegraphics[trim={0cm 0cm 0cm 0cm},clip,width=1\textwidth]{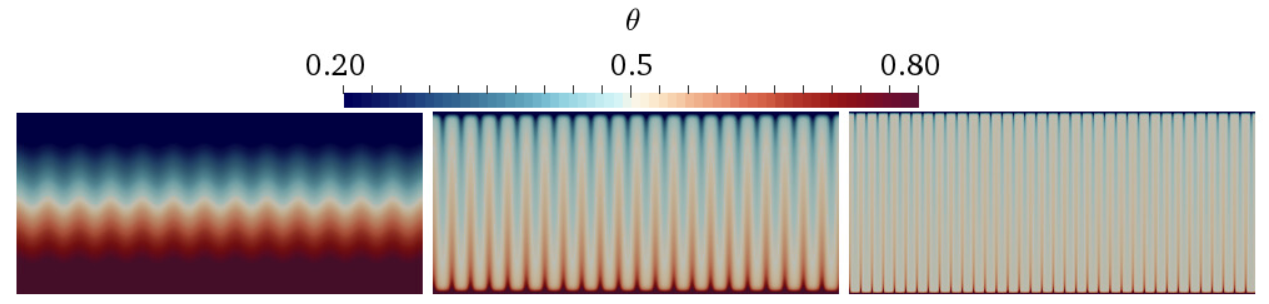} 
    \caption{Visualization of the cellular, columnar and plume regimes at $Q=10^7$ with $Ra =1.1\times 10^8, 8\times 10^8 \text{and } 5\times 10^9  $ respectively. The colormap values have been adjusted to $[0.2,0.8]$ for better visibility.}
    \label{fig:cell_col_plume_regimesQ1e7}
\end{figure}

Figure \ref{fig:nuss_and_lambda_both}a gives a brief overview of the parameter space characterized by dimensionless heat transport. Three distinct scaling regimes with different slopes are observed at large values of $Q$: (1) a cellular regime close to the onset, (2) $Nu \sim Ra$ regime, characterized by the existence of elongated, vertically aligned columns, and (3) $Nu \sim Ra^{1/3}$ regime, which is characterized by the existence of plume-like structures with thin stems and broadened heads. Some select cases representing these regimes are shown in  figure \ref{fig:cell_col_plume_regimesQ1e7}. The corresponding scaling regimes in $(Nu,Ra)$ space are represented by black dashed (columnar) and dotted (plume) lines in figure \ref{fig:nuss_and_lambda_both}a. It is important to note here that the states where transitions occur from cellular to columnar to plume regimes are different in 2D simulations in comparison with 3D simulations. This is because, for 2D simulations, consistent with the mass conservation, we observed a higher $Re_x$ compared to that in 3D at the same $Ra$. To see a comparison of $Re_x$ in 2D and 3D, refer to figure \ref{fig:vertical_profiles_all}e. Higher $Re_x$ is responsible for enhancing the buoyant fluxes in 2D, which ultimately translates into an increased $Nu$, at least for the regime that we are most interested in (the 2D columnar regime), e.g. see figure \ref{fig:nuss_and_lambda_both}a. The differing transition locations are evident. In addition to that, we can also clearly observe the columnar regime in both 2D and 3D, albeit with slightly different $Nu$. The transition locations and the magnitude of the heat flux and flow velocity may differ; however structural similarities exist, and for 2D and 3D alike, the flow is characterized by cellular, columnar, and plume-like structures for the considered $Ra$ range at different values of $Q$. This is important for scaling since a specific regime in 3D is expected and shown to have a counterpart in 2D.

The existence of the columnar and plume regimes can be explained very well by invoking the predictions from marginally stable thermal boundary layer analysis, which predicts $Nu \sim {L}/{\lambda_{\theta}} \sim Ra/Q$ at large values of $Q_{\lambda_{\theta}} = B^2_0 \lambda^2_{\theta}/(\mu\rho\nu\eta) = Q\lambda^2_{\theta}/L^2$, and $Nu \sim {L}/{\lambda_{\theta}} \sim Ra^{1/3}$ at smaller values of $Q_{\lambda_{\theta}}$ due to thinner thermal boundary layer thickness in the plume regime \citep{bhattacharjee1991turbulent}. See figure \ref{fig:nuss_and_lambda_both}b.

In the following sections, for the columnar regime, we will explain the scaling laws concerning not only the heat transport, but also the horizontal length scale, flow velocity, and Ohmic dissipation by using energetic arguments. To begin, we start with the exact relations for time- and volume-averaged (area-averaged in 2D) viscous dissipation rate $\epsilon_{\nu} = \nu \langle  \overline{|\nabla \mathbf{u}|^2} \rangle$, Ohmic dissipation rate $\epsilon_{\eta} = (\rho \sigma)^{-1} \langle  \overline{|\mathbf{J}|^2} \rangle $, and the thermal dissipation rate $\epsilon_{\theta} = \kappa \langle \overline{|\nabla \theta|^2} \rangle $ \textcolor{black}{\citep{grossmann2000scaling, RevModPhys.81.503, bhattacharyya2006scaling,song2023scaling}},

\begin{align}
    \epsilon_{\nu} + \epsilon_{\eta} &= \frac{\nu^3}{L^4} (Nu-1)RaPr^{-2} \label{eqn:exactDiss_v_o} \\
    \epsilon_{\theta} &= \kappa \frac{\Delta^2}{L^2}Nu \label{eqn:exactDiss_k}
\end{align}

The convective part of $\epsilon_{\theta}$ can be expressed as,

\begin{align}
    \Tilde{\epsilon}_{\theta} &= \epsilon_{\theta} - \epsilon_{cond} 
    \implies Nu-1 \sim  \frac{\Tilde{\epsilon}_{\theta}}{\kappa \Delta^2/L^2} \label{eqn:nu_convc_dissk} 
\end{align}

Introducing $u, \theta \ \text{and} \ \ell$ as the representative scales for convective velocity, temperature, and the length scale, the convective part of thermal dissipation rate can be expressed as $ \Tilde{\epsilon}_{\theta} \sim u\theta^2/\ell$. Using this, after rearranging, we can rewrite equation \ref{eqn:nu_convc_dissk} as \citep{song2023scaling},

\begin{align}\label{eqn:nu_repr}
    Nu-1 \sim \frac{\theta^2}{\Delta^2} \frac{L}{\ell} Re Pr
\end{align}

$Nu$ can also be expressed in terms of convective heat transfer $q\sim u\theta$, as,

\begin{align}\label{eqn:nu_from_q}
    Nu-1 \sim \frac{u \theta }{\kappa \Delta/L}
\end{align}

From equations \ref{eqn:nu_repr} and \ref{eqn:nu_from_q}, we get $\ell/L \sim \theta/\Delta$, leading to,

\begin{align}\label{eqn:Nu_lLRePr}
    Nu-1 \sim \left ( \frac{\ell}{L} \right) Re Pr
\end{align}

Assuming the induction $\mathbf{b}$, the total field can be written as $\mathbf{B} = \mathbf{B_0} + \mathbf{b}$. From Ohmic dissipation $\epsilon_{\eta} = (\rho \sigma)^{-1} \langle  \overline{|\mathbf{J}|^2} \rangle$, and $\mathbf{J} = (\nabla \times \mathbf{B})/\mu$, and assuming the length scale is $\ell$ we can write,

\begin{align}\label{eqn:ohmDissipeEstimate}
    \epsilon_{\eta} = \frac{1}{\rho \sigma \mu^2} \langle  \overline{|\nabla \times \mathbf{B}|^2}  \rangle \sim \frac{1}{\rho \sigma \mu^2} \frac{b^2}{\ell^2},
\end{align}
where $b$ represents the typical magnitude of the induced field.
Under the quasi-static assumption, the induction equation can be simplified to,

\begin{align}
    \eta \nabla^2 \mathbf{b} + B_0   \frac{\partial {u_x}}{\partial x} &= 0
\end{align}

which upon using the scale estimates for $\mathbf{b} \sim b$, ${u_x
} \sim u$, $\partial^2/\partial x^2 \sim 1/L^2$, $\partial^2/\partial y^2 \sim 1/\ell^2$, $1/\ell^2 >> 1/L^2$ and $\partial/\partial x \sim 1/L$ becomes,

\begin{align}\label{eqn:bEstimate}
     \eta \frac{b}{\ell^2} &\sim B_0 \frac{u}{L} 
     \implies b \sim \frac{B_0}{\eta} \frac{u \ell^2}{L}
\end{align}

Using this in equation \ref{eqn:ohmDissipeEstimate}, after rearranging and grouping relevant terms, we finally obtain a scaling law for the Ohmic dissipation with the horizontal length scale dependence,

\begin{align}\label{eqn:ohmicdiss_Re_Q_l}
    \epsilon_{\eta} \sim \frac{\nu^3}{L^4} Re^2 Q \left (  \frac{\ell}{L}  \right )^2
\end{align}

\begin{figure}
\captionsetup{skip=0pt}
     \centering
     \begin{subfigure}[b]{0.45\textwidth}
         \centering
         \includegraphics[width=1\textwidth]{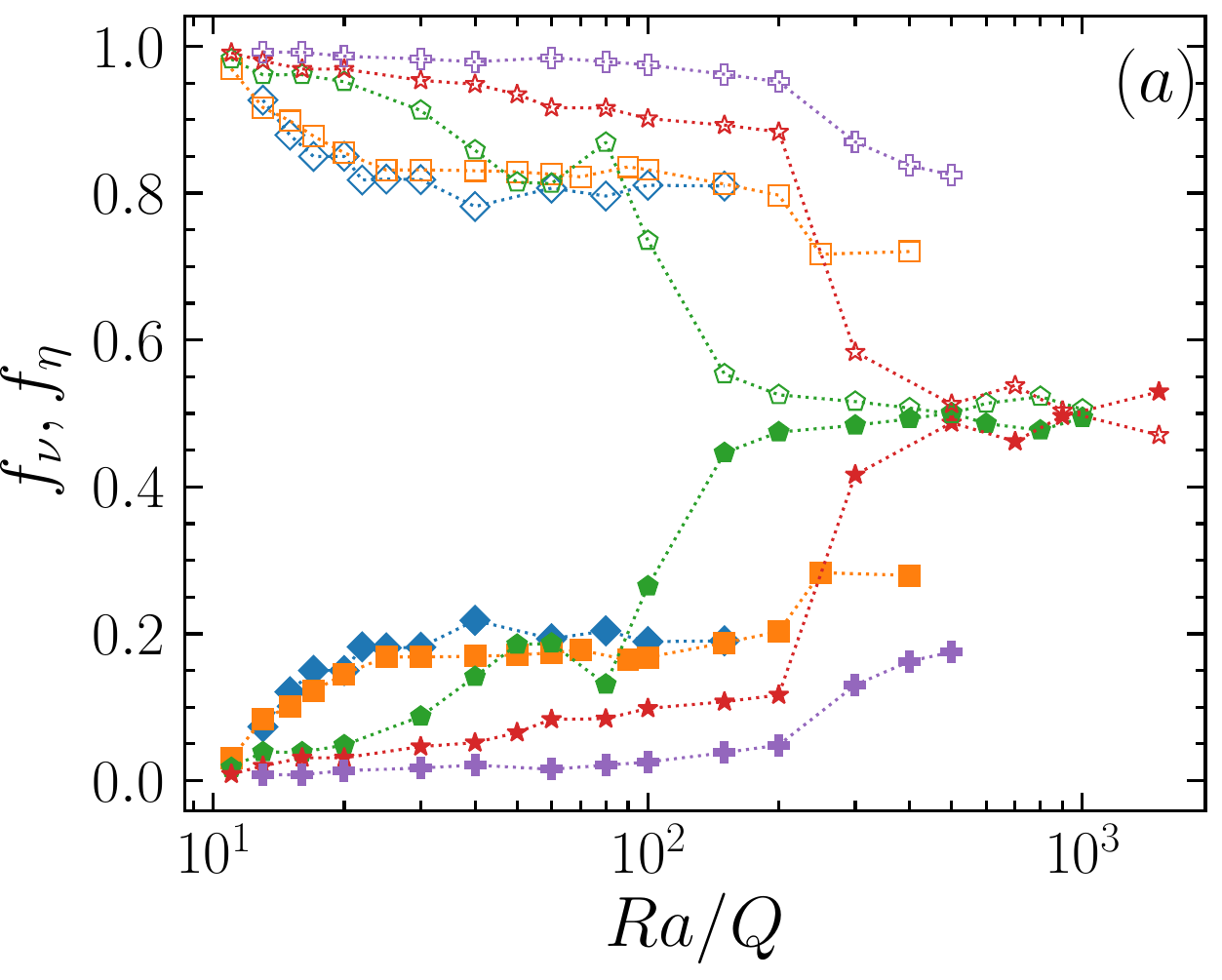}
     \end{subfigure}
     \hspace{0.1cm}
     \begin{subfigure}[b]{0.45\textwidth}
         \centering
         \includegraphics[width=1\textwidth]{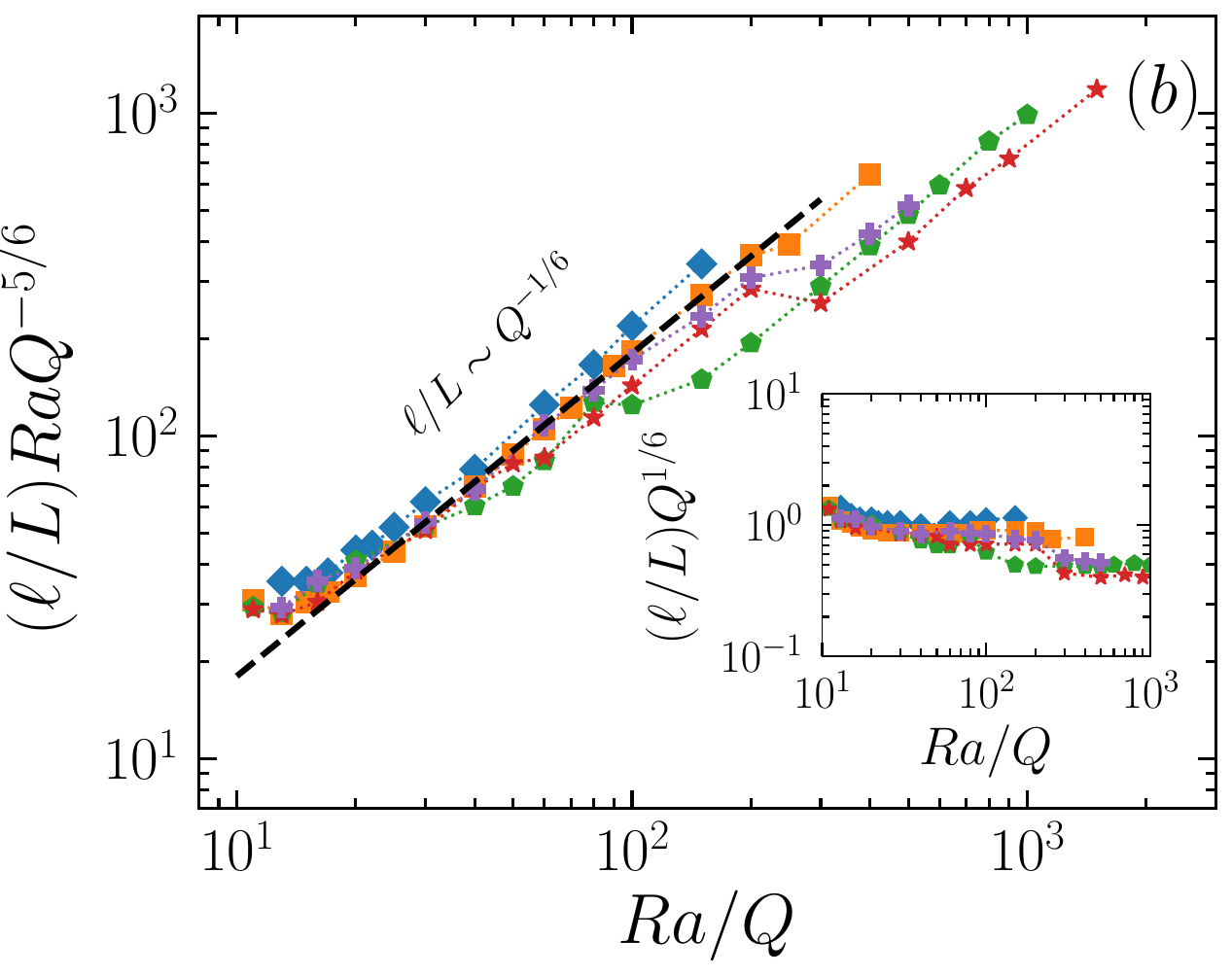}
     \end{subfigure}
        \caption{(a) Fraction of viscous dissipation $f_{\nu} = \epsilon_{\nu}/(\epsilon_{\nu} + \epsilon_{\eta})$ (solid symbols) and Ohmic dissipation $f_{\eta} = \epsilon_{\eta}/(\epsilon_{\nu} + \epsilon_{\eta})$ (open symbols), (b) Dependence of the horizontal length scale on $Q$ in the columnar regime. The color for both the open and solid symbols is the same as in figure \ref{fig:nuss_and_lambda_both}.}
        \label{fig:frac_n_ell}
\end{figure}

With $Re$ and $\ell$ dependence of $Nu$ and $\epsilon_{\eta}$, we will proceed to derive the scaling relation of $Re \sim Re(Ra,Q,Pr,\ell/L)$ and $\ell/L \sim Q^{\gamma}$ in the columnar regime. It is to be noted the length scale of the structures in the columnar regime varies negligibly with $Ra$, which is why its $Ra$ dependency is not considered. In figure \ref{fig:frac_n_ell}a, in the columnar regime, we observe $\epsilon_{\eta} >> \epsilon_{\nu}$. Hence from the equation \ref{eqn:exactDiss_v_o}, we can consider $\epsilon_{\eta} \approx (\nu^3/L^4) (Nu-1)RaPr^{-2}$. Combining this together with equation \ref{eqn:ohmicdiss_Re_Q_l} and \ref{eqn:Nu_lLRePr}, we derive the scaling law for $Re$ in terms of the input parameters and the horizontal length scale,

\begin{align}\label{eqn:Re_QRaPrlL}
    Re \sim Ra Q^{-1}  Pr^{-1}\left (  \frac{\ell}{L}  \right
 )^{-1}
\end{align}

Using this in equation \ref{eqn:Nu_lLRePr}, length scale and Prandtl number dependence drops out, yielding the Nusselt number scaling,

\begin{align}\label{eqn:Nu_law}
    Nu -1 \sim \frac{Ra}{Q}
\end{align}

\begin{figure}
\captionsetup{skip=0pt}
     \centering
     \begin{subfigure}[b]{0.45\textwidth}
         \centering
         \includegraphics[width=\textwidth]{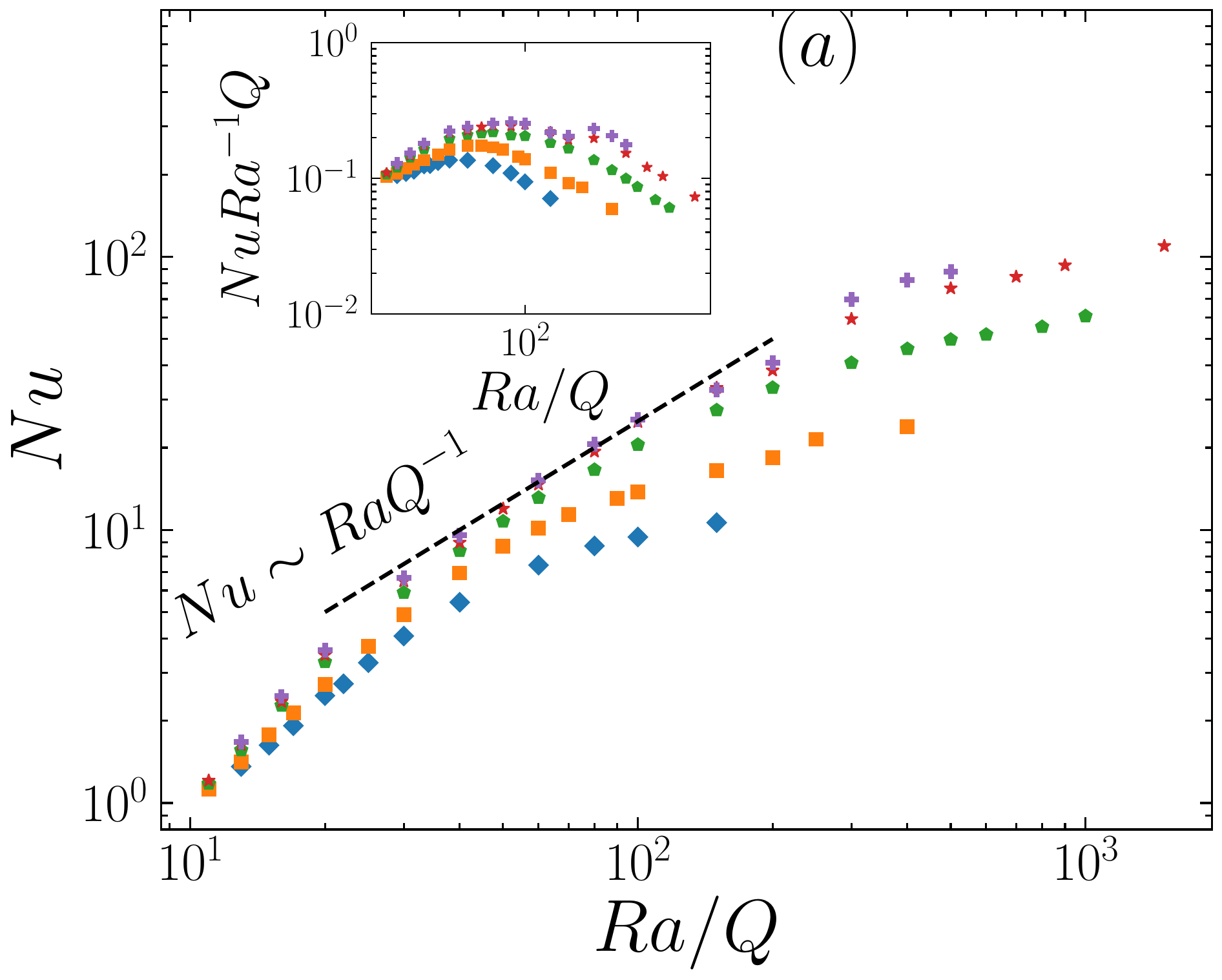}
     \end{subfigure}
     \begin{subfigure}[b]{0.45\textwidth}
         \centering
         \includegraphics[width=\textwidth]{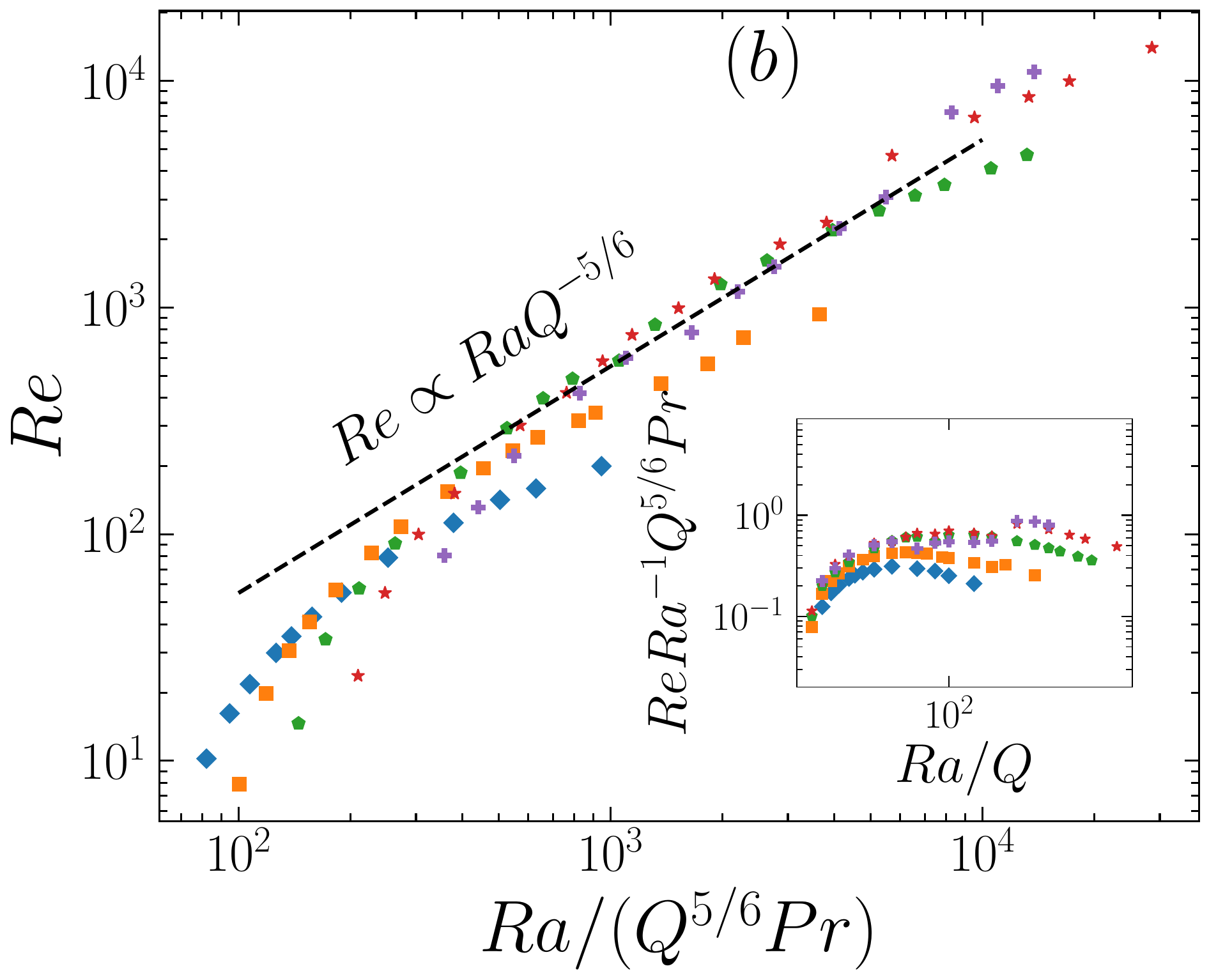}
     \end{subfigure}
     \\ \vspace{0.25cm}
     \begin{subfigure}[b]{0.45\textwidth}
         \centering
         \includegraphics[width=\textwidth]{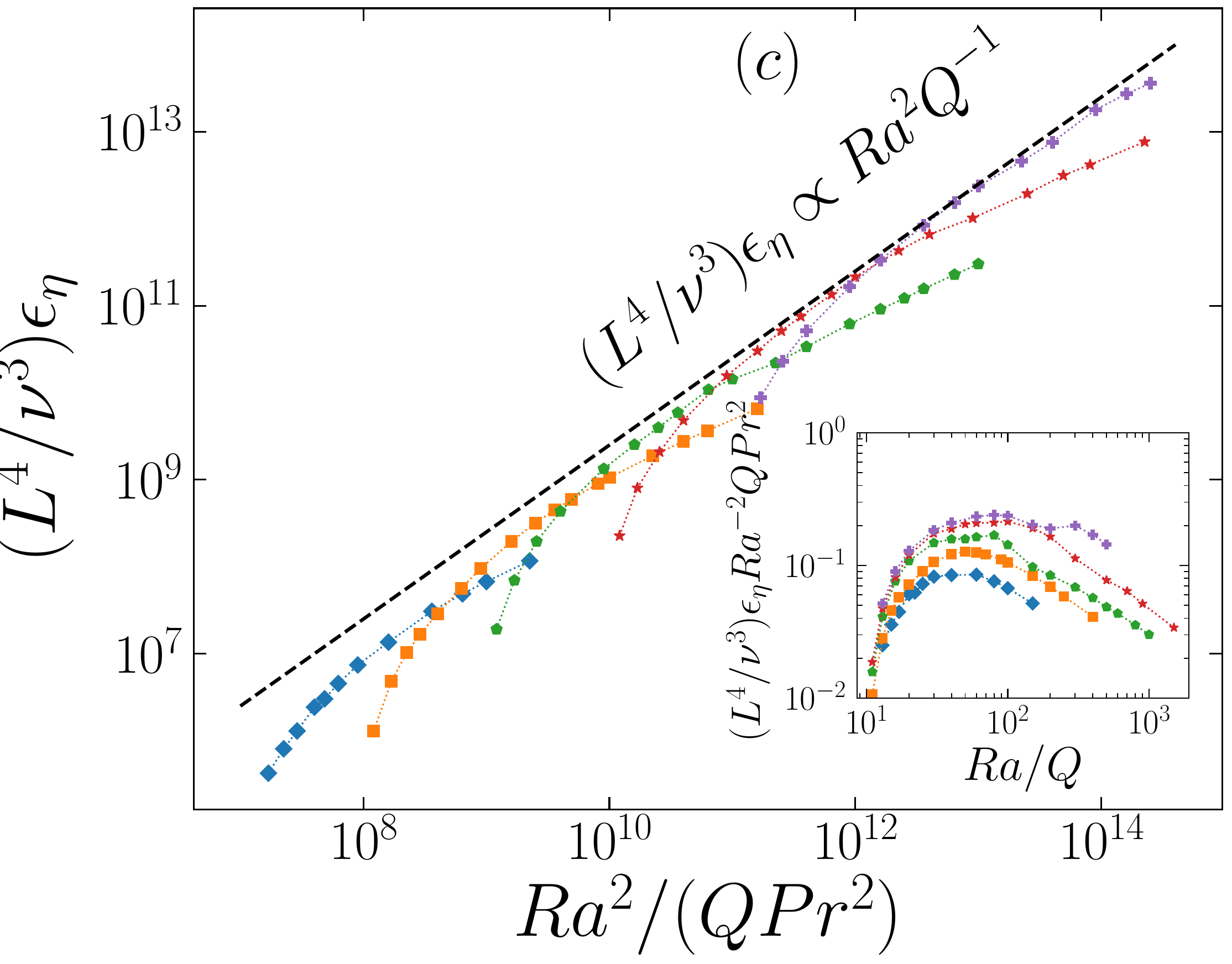}
     \end{subfigure}
        \caption{ Scaling of the (a) Nusselt number, (b) Reynolds number, and (c) Ohmic dissipation in the columnar regime.}
        \label{fig:NuReyOhmicdiss_all}
\end{figure}

This result is consistent with the scaling law obtained from the marginally stable thermal boundary layer analysis discussed above. For non-magnetic convection, a regime in which the entire fluid layer is turbulent (at asymptotically large values of $Ra$) is referred to as the ultimate regime. This regime, predicted by \citet{kraichnan1962turbulent}, is characterized by the heat transport becoming independent of the diffusion coefficients. With additional constraining forces, like the Lorentz force in the present case, a heat transport regime independent of the diffusion coefficients $(\nu \text{ and } \kappa)$ can be observed at relatively low Rayleigh numbers. It is noteworthy to point out that in the columnar regime ($Nu-1 \sim Ra/Q$), an inherently different flow state in comparison to the ultimate state of conventional RBC, the heat transport $Nu \times \kappa \Delta /L$ in the limit of an asymptotically large $Q$, is independent of $\nu$ and $\kappa$. It does, however, depend on the magnetic diffusion coefficient $\eta$. It is also evident that the scaling of the length scale has no bearing upon the scaling law for the dimensionless convective heat transport in the columnar regime. Equation \ref{eqn:Nu_law} is validated against our data in figure \ref{fig:NuReyOhmicdiss_all}a.

One way to obtain the dependence of the horizontal length scale on the Chandrasekhar number is to obtain the scaling for the critical horizontal wavenumber $k_c$. From linear theory, it is well known that at large $Q$, $k_c \sim Q^{1/6}$, which implies $l_c \sim Q^{-1/6}$ \citep{chandra}. In figure \ref{fig:frac_n_ell}b, the $Q$-dependence of the length scale found in the data is compared to this result from the linear theory. We find a good collapse of data for all the simulation sets according to $\ell/L \sim Q^{-1/6}$ in the columnar regime. Using this in equations \ref{eqn:ohmicdiss_Re_Q_l} and \ref{eqn:Re_QRaPrlL}, we obtain the following relations for the Reynolds number and Ohmic dissipation scaling,

\begin{align}
    Re &\sim RaQ^{-5/6} Pr^{-1} \label{eqn:Rescalingfinal}\\
    \epsilon_{\eta}\frac{L^4}{\nu^3} &\sim Re^2 Q^{2/3} \sim Ra^2 Q^{-1} Pr^{-2} \label{eqn:ohmicscalingfinal}
\end{align}

Data from our simulations show a good collapse for the above relations. See figure \ref{fig:NuReyOhmicdiss_all}. In figure \ref{fig:NuReyOhmicdiss_all}a, at large $Q$, the $Nu$ data collapse in the columnar regime. Similarly, in figure \ref{fig:NuReyOhmicdiss_all}b, the $Re$ data at large $Q$ collapse, validating equation \ref{eqn:Rescalingfinal}. \textcolor{black}{It must however be noted for Reynolds number scaling, the exponent differs from one suggested by \citet{yan_calkins_maffei_julien_tobias_marti_2019} for 3D quasi-static magnetoconvection. They derive the scaling relation for $Re$ by balancing the Lorentz force with the buoyancy in the columnar regime, which is evident from their 3D data. Our derivations are based on energetic arguments involving the dissipation rates. Different assumptions invoked in these arguments is expected to lead to the differences observed in the exponent.}  The Ohmic dissipation data, shown in figure \ref{fig:NuReyOhmicdiss_all}c, at all $Q$ show a solid trend towards the scaling given by equation \ref{eqn:ohmicscalingfinal}, collapsing for $Q$ in the asymptotic limit, $Q\rightarrow \infty$, in the columnar regime. 

\section{Concluding remarks}
The effect of the horizontal length scale on the scaling behavior of the response parameters in quasi-static magnetoconvection has not been explored in the literature. Previous studies concerning the energetic scaling arguments have considered the domain height $L$ as the sole length scale present in the system. In the present study, under the quasi-static assumption, we consider the horizontal width of the columns as a relevant length scale and explore the effect of its dependence on the Chandrasekhar number on the scaling of the dimensionless heat transport, flow velocity, and Ohmic dissipation. Based on our DNS data, the derived scaling laws are validated successfully for Nusselt number, Reynolds number, and Ohmic dissipation. The Nusselt number scaling obtained from our analysis is shown to be consistent with the marginal stability predictions, which further validates our theoretical analysis. The scaling law for the Reynolds number and Ohmic dissipation also shows a reasonable collapse in our data at large $Q$, implying the validity of our scaling arguments. 
\par
\textcolor{black}{In the present work, we have considered periodicity in the horizontal direction and a fixed Prandtl number of unity. Compared to periodic boundaries, the presence of sidewalls has been shown to lead to convection at a much lower $Ra$ \citep{busse2008asymptotic}. Numerical simulations of planar RBC by \citet{liu2018wall} confirmed the existence of wall-modes at $Ra \lesssim Ra_c$. \citet{akhmedagaev2020turbulent} also confirmed the existence of these wall-modes at $Ra \gtrsim Ra_c$. The effect of sidewalls at even higher $Q$ and $Ra$ away from the onset will be something that we will explore in the future editions of this work. The presence of sidewalls may have implications in the design of liquid metal batteries \citep{liquidMetalBattery} and nuclear fusion technology \citep{fusionTech}, wherein the geometries are often confined by walls.}
\par 
\textcolor{black}{Magnetoconvection is usually studied in liquid metals, which have very small Prandtl numbers and very high electrical conductivities.  
In the numerical and experimental low $Pr$ studies of \citet{akhmedagaev2020turbulent} and \citet{zurner2020flow}, unlike the slender columns and magneto-plumes observed in our simulations, the dominant structures were observed to be a system of ascending and descending planar jets originating at the sidewalls. It must be noted, however, that the aforementioned studies consider convection in a cylindrical container in which sidewalls play a dominant role in organizing the flow. Hence, it remains to be seen if the slender columnar or plume structures can emerge in 2D simulations by virtue of a low Prandtl number or exclusively in the presence of sidewalls. Whether or not the flow structures are different, we believe that it does not affect the scaling argument for the conditions that we have considered. That can be seen from the low $Pr$ ($Pr=0.025$) simulations of \citet{yan_calkins_maffei_julien_tobias_marti_2019}, who performed the comparative calculations at $Pr = 1$ and $Pr = 0.025$. In their work, it was observed that the scaling behavior of the dimensionless heat transport seems to follow the same trend as in the case of $Pr$ being unity, i.e. $Nu \sim Ra/Q$ at asymptotically large values of $Q$; since in the columnar regime, there is no $Pr$ dependence of $Nu$. Although, $Re$ and $\epsilon_{\eta}$ show $Pr$ dependence in this regime, in this work, our main goal was to test the scaling argument that we proposed for aymptotically large $Q$. $Re$ and $\epsilon_{\eta}$ scaling relations show a good agreement with what we observed in our DNS data (for $Q$ and $Ra$ dependence). Exploration of the $Pr$ dependence of the response parameters and the flow structures is a full-fledged campaign in itself that we propose to study in future.}

\vspace{1cm} \par
{\small\textbf{Acknowledgements} We thank Jiaxing Song for helpful discussions. We gratefully acknowledge the support from the Max Planck Society, the Alexander von Humboldt Foundation, and German Research Foundation (DFG), and the computing time provided 
on the high-performance computer Lichtenberg at the NHR Centers NHR4CES,
 at TU Darmstadt, funded by the Federal Ministry of Education and Research, and the state governments participating on the basis of the resolutions of the GWK for national high-performance computing at universities, on 
the HPC systems of Max Planck Computing and Data Facility (MPCDF), 
the HoreKa supercomputer, funded by the Ministry of Science, Research and the Arts Baden-W\"urttemberg and by the Federal Ministry of Education and Research, 
and the GCS Supercomputer SuperMUC-NG at Leibniz Supercomputing Centre.}

\vspace{0.5cm}  \par 
{\small\textbf{Declaration of Interests}. The authors report no conflict of interest.}

\appendix
\section{Appendix}\label{appA}
To ensure that the flow is well resolved, the resolution is chosen to be able to capture the Kolmogorov length scales based on both the viscous as well as Ohmic dissipation. The horizontal size of the 2D domain is selected to resolve at least $10\lambda_c$, where $\lambda_c$, which is known from the linear theory at a given $Q$, is the critical horizontal wavelength. Table \ref{tab:data} below lists the simulation data. 
\begin{longtable}{r@{\hspace{1cm}} r@{\hspace{1cm}} r@{\hspace{1cm}} r@{\hspace{1cm}} r}
\hline
  $Ra$ &  $Re$ &  $Nu_{m}$ &  $Error (\%)$ &   $N_x \times N_y $ \\ 
\hline \\[-1.5ex]
   &      & $Q=10^5$, $\Gamma=5$ &        &          \\
  \convert{1300000.0} &   10.22 &    1.36 &   0.07 &  $48 \times 512$ \\
  \convert{1500000.0} &   16.18 &    1.63 &   0.12 &  $48 \times 512$ \\
  \convert{1700000.0} &   21.80 &    1.92 &   0.19 &  $48 \times 512$ \\
  \convert{2000000.0} &   29.94 &    2.47 &   0.31 &  $48 \times 512$ \\
  \convert{2200000.0} &   35.40 &    2.73 &   0.38 &  $48 \times 512$ \\
  \convert{2500000.0} &   43.21 &    3.26 &   0.25 &  $64 \times 768$ \\
  \convert{3000000.0} &   55.21 &    4.09 &   0.35 &  $64 \times 768$ \\
  \convert{4000000.0} &   78.78 &    5.43 &   0.52 &  $64 \times 768$ \\
  \convert{6000000.0} &  112.24 &    7.43 &   0.27 &  $96 \times 1024$ \\
  \convert{8000000.0} &  141.76 &    8.74 &   0.33 &  $96 \times 1024$ \\
  \convert{10000000.0} &  159.08 &    9.42 &   0.35 &  $96 \times 1024$ \\
  \convert{15000000.0} &  199.59 &   10.64 &   0.08 &  $192 \times 2048$ \\
\hline \\[-1.5ex]
     &      & $Q=10^6$, $\Gamma=3.5
$&        &          \\
   \convert{11000000.0} &    7.87 &    1.12 &   0.02 &  $48 \times 384$ \\
   \convert{13000000.0} &   19.78 &    1.41 &   0.08 &  $48 \times 384$ \\
   \convert{15000000.0} &   30.57 &    1.78 &   0.09 &  $64 \times 512$ \\
   \convert{17000000.0} &   41.08 &    2.14 &   0.13 &  $64 \times 512$ \\
   \convert{20000000.0} &   56.75 &    2.72 &   0.21 &  $64 \times 512$ \\
   \convert{25000000.0} &   82.90 &    3.75 &   0.13 &  $96 \times 768$ \\
   \convert{30000000.0} &  107.79 &    4.88 &   0.20 &  $96 \times 768$ \\
   \convert{40000000.0} &  153.75 &    6.95 &   0.31 &  $96 \times 768$ \\
   \convert{50000000.0} &  195.35 &    8.70 &   0.07 &  $192 \times 1152$ \\
   \convert{60000000.0} &  233.00 &   10.15 &   0.09 &  $192 \times 1152$ \\
   \convert{70000000.0} &  267.63 &   11.37 &   0.10 &  $192 \times 1152$ \\
   \convert{90000000.0} &  316.38 &   13.03 &   0.11 &  $192 \times 1152$ \\
  \convert{100000000.0} &  343.53 &   13.78 &   0.12 &  $192 \times 1152$ \\
  \convert{150000000.0} &  462.75 &   16.52 &   0.08 &  $256 \times 1536$ \\
  \convert{200000000.0} &  564.22 &   18.39 &   0.09 &  $256 \times 1536$ \\
  \convert{250000000.0} &  738.49 &   21.46 &   0.11 &  $256 \times 1536$ \\
  \convert{400000000.0} &  933.23 &   23.82 &   0.14 &  $256 \times 1536$ \\
 \convert{800000000.0} &  1750.32 &   36.74 &   0.05 &  $512 \times 3072$ \\
\hline \\[-1.5ex]
     &      & $Q=10^7$, $\Gamma=2.25$&        &          \\
 \num{1.1e+08} &    14.62 &    1.18 &   0.01 &  $96 \times 384$ \\
 \num{1.3e+08} &    34.33 &    1.56 &   0.03 &  $96 \times 384$ \\
 \num{1.6e+08} &    57.55 &    2.28 &   0.03 &  $128 \times 576$ \\
 \num{2.0e+08} &    90.89 &    3.28 &   0.02 &  $192 \times 768$ \\
 \num{3.0e+08} &   186.58 &    5.89 &   0.05 &  $192 \times 768$ \\
 \num{4.0e+08} &   293.38 &    8.38 &   0.04 &  $256 \times 864$ \\
 \num{5.0e+08} &   397.24 &   10.74 &   0.05 &  $256 \times 864$ \\
 \num{6.0e+08} &   484.78 &   13.12 &   0.06 &  $288 \times 1152$ \\
 \num{8.0e+08} &   582.52 &   16.59 &   0.04 &  $384 \times 1536$ \\
 \num{1.0e+09} &   839.61 &   20.50 &   0.05 &  $384 \times 1536$ \\
 \num{1.5e+09} &  1266.29 &   27.42 &   0.03 &  $576 \times 2048$ \\
 \num{2.0e+09} &  1611.88 &   33.14 &   0.02 &  $768 \times 2592$ \\
 \num{3.0e+09} &  2192.99 &   40.87 &   0.03 &  $768 \times 2592$ \\
 \num{4.0e+09} &  2683.44 &   45.96 &   0.02 &  $864 \times 3072$ \\
 \num{5.0e+09} &  3117.13 &   49.72 &   0.02 &  $1024 \times 3456$ \\
 \num{6.0e+09} &  3474.82 &   51.87 &   0.02 &  $1024 \times 3456$ \\
 \num{8.0e+09} &  4108.87 &   55.36 &   0.02 &  $1024 \times 3456$ \\
 \num{1.0e+10} &  4712.78 &   60.63 &   0.02 &  $1152 \times 4096$ \\
\hline  \\[-1.5ex]
     &      & $Q=10^8$, $\Gamma=1.5$&        &          \\
 \num{1.1e+09} &     23.72 &    1.21 &   0.00 &  $192 \times 512$ \\
 \num{1.3e+09} &     55.02 &    1.63 &   0.01 &  $192 \times 512$ \\
 \num{1.6e+09} &     99.74 &    2.35 &   0.01 &  $256 \times 576$ \\
 \num{2.0e+09} &    151.13 &    3.48 &   0.01 &  $384 \times 864$ \\
 \num{3.0e+09} &    302.00 &    6.49 &   0.06 &  $512 \times 1296$ \\
 \num{4.0e+09} &    420.63 &    8.99 &   0.31 &  $512 \times 1296$ \\
 \num{5.0e+09} &    579.38 &   11.96 &   0.44 &  $512 \times 1152$ \\
 \num{6.0e+09} &    757.51 &   14.66 &   0.51 &  $576 \times 1296$ \\
 \num{8.0e+09} &    993.48 &   19.36 &   0.23 &  $768 \times 1728$ \\
 \num{1.0e+10} &   1334.82 &   24.84 &   0.11 &  $768 \times 1728$ \\
 \num{1.5e+10} &   1901.78 &   33.12 &   0.01 &  $1024 \times 2560$ \\
 \num{2.0e+10} &   2370.53 &   38.35 &   0.01 &  $1024 \times 2560$ \\
 \num{3.0e+10} &   4679.94 &   59.18 &   0.01 &  $1536 \times 3456$ \\
 \num{5.0e+10} &   6896.39 &   76.64 &   0.01 &  $2048 \times 4608$ \\
 \num{7.0e+10} &   8498.61 &   84.48 &   0.01 &  $2048 \times 4608$ \\
 \num{9.0e+10} &   9979.42 &   93.08 &   0.02 &  $2560 \times 5760$ \\
 \num{1.5e+11} &  14009.06 &  109.52 &   0.02 &  $2560 \times 5760$ \\

\hline \\[-1.5ex]
     &      & $Q=10^9$, $\Gamma=1$&        &          \\
 \num{1.3e+10} &     80.37 &    1.68 &   0.01 &  $384 \times 576$ \\
 \num{1.6e+10} &    131.44 &    2.46 &   0.04 &  $512 \times 768$ \\
 \num{2.0e+10} &    221.78 &    3.62 &   0.24 &  $512 \times 768$ \\
 \num{3.0e+10} &    419.29 &    6.67 &   1.25 &  $768 \times 1152$ \\
 \num{4.0e+10} &    600.36 &    9.56 &   0.37 &  $864 \times 1296$ \\
 \num{6.0e+10} &    776.00 &   15.22 &   0.28 &  $1152 \times 1728$ \\
 \num{8.0e+10} &   1173.95 &   20.65 &   0.05 &  $1296 \times 2048$ \\
 \num{1.0e+11} &   1509.97 &   25.35 &   0.04 &  $1536 \times 2048$ \\
 \num{1.5e+11} &   2236.29 &   32.61 &   0.01 &  $1536 \times 2048$ \\
 \num{2.0e+11} &   3062.54 &   40.90 &   0.01 &  $2048 \times 3072$ \\
 \num{3.0e+11} &   7260.48 &   69.85 &   0.00 &  $2560 \times 3840$ \\
 \num{4.0e+11} &   9529.59 &   82.35 &   0.01 &  $2560 \times 3840$ \\
 \num{5.0e+11} &  10958.72 &   88.36 &   0.04 &  $3072 \times 4608$ \\
 \hline \hline
 \caption{Simulation details. $Q$ is the Chandrasekhar number, $Ra$ is the Rayleigh number, $Nu_{m}$ is the mean of the Nusselt numbers computed using the thermal dissipation, kinetic energy dissipation via viscous and Ohmic counterparts, plate Nusselt number, and the Nusselt number based on volume-averaged heat flux. Error represents the maximum deviation (in percentage) between the smallest and the largest individual Nusselt numbers. $Re$ is the Reynolds number, $\Gamma$ is the domain aspect ratio, $N_x \text{ and } N_y$ represent the resolution of the grid in vertical and horizontal directions.}
 \label{tab:data}
\end{longtable}

\bibliographystyle{unsrtnat}
\bibliography{ms}

\end{document}